# Time-encoded pseudo-continuous arterial spin labeling: increasing SNR in ASL angiography

Joseph G. Woods[1], S. Sophie Schauman,[1,2] Mark Chiew,[1] Michael A. Chappell,[1,3,4,5] Thomas W. Okell[1]

[1]Wellcome Centre for Integrated Neuroimaging, FMRIB, Nuffield Department of Clinical Neuroscience, University of Oxford, Oxford, UK

[2]Department of Radiology, Stanford University, Stanford, California, USA

[3]Mental Health and Clinical Neurosciences, School of Medicine, University of Nottingham, Nottingham, UK

[4]Sir Peter Mansfield Imaging Centre, School of Medicine, University of Nottingham, Nottingham, UK

[5]Nottingham Biomedical Research Centre, Queen's Medical Centre, University of Nottingham, Nottingham, UK

**Correspondence**

Joseph G. Woods, Wellcome Centre for Integrative Neuroimaging, FMRIB Building, John Radcliffe Hospital, Headington, Oxford, OX3 9DU

Email: joseph.woods@ndcn.ox.ac.uk

Abstract word count: 250

Manuscript word count: 5281

Figure count: 9

Table count: 0

## Abstract:

**Purpose:** Dynamic angiography using arterial spin labeling (ASL) can provide detailed hemodynamic information. However, the long time-resolved readouts require small flip angles to preserve ASL signal for later timepoints, limiting SNR. By using time-encoded ASL to generate temporal information, the readout can be shortened. Here, the improvements in SNR from using larger flip angles, made possible by the shorter readout, are quantitatively investigated.

**Methods:** The SNR of a conventional (sequential) protocol with 9 readouts and a 4-by-3 time-encoded protocol with 3 readouts (giving 9 matched timepoints) were directly compared using simulations and in vivo data. Both protocols were compared using readouts with constant (CFA) and variable flip angles (VFA), where the VFA scheme was designed to produce a consistent ASL signal across readouts. Optimization of the background suppression to minimize physiological noise across readouts was also explored.

**Results:** The time-encoded protocol increased in vivo SNR by 103% and 96% when using CFAs or VFAs, respectively. Use of VFAs improved SNR compared to CFAs by 25% and 21% for the sequential and time-encoded protocols, respectively. The VFA scheme also removed signal discontinuities in the time-encoded data. Preliminary data suggest optimizing the background suppression could improve in vivo SNR by a further 16%.

**Conclusion:** Time-encoding can be used to generate additional temporal information in ASL angiography. This enables the use of larger flip angles which can double the SNR compared to a non-time-encoded protocol. The shortened time-encoded readout can also lead to improved background suppression, reducing physiological noise and further improving SNR.

# 1. Introduction

Dynamic angiograms provide much richer hemodynamic[1,2] information than static angiograms and can be acquired with an arterial spin labeling (ASL) preparation and highly segmented time-resolved (Look-Locker)[3] readouts.[4–6] However, due to the large number of excitations, low flip-angles are necessary to preserve signal for later timepoints, limiting SNR.[7] By encoding some, or all, of the desired temporal information into the PCASL pulse train using a time-encoded preparation,[8] higher flip-angles can be used because fewer excitations are required to achieve a given temporal resolution (see Theory and Figure 1).[9,10]

While the SNR benefits of time-encoded PCASL have been demonstrated for perfusion imaging,[11–13] they have not been quantitatively investigated for dynamic angiography. The SNR in ASL-based perfusion imaging is dependent on the accumulation of labeled spins in the tissue, so the shorter label durations (LDs) required with time-encoded PCASL compared to sequential PCASL can lead to a reduction in the overall SNR benefit of the technique for perfusion imaging.[13] In contrast, ASL angiography is not dependent on tissue signal accumulation because labeled blood is visualized as it traverses the arterial vasculature, meaning that short LDs do not decrease the peak signal (unless the LDs approach the timescale of bolus dispersion).

Here, we investigated the SNR benefits from using time-encoded PCASL to generate temporal information for dynamic angiography, enabling larger flip angles to be used during a spoiled gradient-echo (SPGR) readout. We also demonstrate how a variable flip angle (VFA) scheme can further improve SNR and remove signal discontinuities between temporal data decoded from different time-encoded blocks. However, because time-encoding increases the minimum scan time, we demonstrate that SNR benefits remain after under-sampling the time-encoded readout to match the scan duration of a fully-sampled sequential scan. Finally, we explore how shortening the readout allows better suppression of background tissue signal during the readout, thus reducing physiological noise and further improving SNR.

# 2. Theory

Time-encoding enables multiple PLDs to be encoded into the PCASL preparation rather than using a time-resolved readout[8] or requiring the PLD/LD to be varied across ASL preparations.[14] To achieve this, the PCASL preparation is split up into multiple distinct time-encoded blocks. The label and control conditions for each block are varied independently over multiple ASL preparations using a Hadamard encoding scheme so that when the control-label difference is decoded for a particular block, the ASL signal contributions from the other blocks cancel out (see Figure 3 in van Osch et al.[15] for a detailed illustration). By using a 4-by-3 Hadamard encoding scheme with a single readout frame, 3 separate PLDs can be decoded. By pairing this encoding with 3 readout frames, a total of 9 separate PLDs can be decoded, as demonstrated in Figure 1. Compared to a sequential preparation, where all the PLDs are directly acquired from the readout frames, the readout duration and number of excitations are reduced by a factor of 1/3.

Reducing the number of excitations with time-encoding enables larger flip angles to be used throughout the SPGR readout. However, when constant flip angles (CFAs) are used, there can be large signal discontinuities between timepoints decoded from different encoded blocks. This is because the data decoded from each time-encoded block from later excitations has been cumulatively attenuated by the preceding excitations, while the data decoded from the earliest excitations have been minimally attenuated. To alleviate this problem, a variable flip angle (VFA) scheme can be used such that the flip angle increases during the readout to exactly compensate for the attenuation of the preceding excitation pulses, resulting in a consistent ASL signal over time.[10] For a SPGR readout, where the $n^{th}$ excitation pulse attenuates the ASL signal by a factor of $\cos(\theta(n))$ and $\theta(n)$ is the $n^{th}$ flip angle, the formula for such a VFA scheme can be defined using a backwards recursive formula for a given maximum flip angle at the last excitation:

$$\theta(n) = \tan^{-1}(\sin(\theta(n+1))), \qquad [1]$$

where $n \in [1, \ldots, N-1]$ and N is the total number of excitations in the readout. This is similar to the formula given by Wang et al.,[6] except that $T_1$-relaxation is not included here because, ignoring dispersion, this will be constant for PCASL signal reaching a given voxel, unlike for PASL. The acquired ASL signal using this equation is maximized when $\theta(N) = 90°$.

## 3. Methods

### 3.1. Protocols

We compared a sequential PCASL protocol with 9 readout frames to a time-encoded PCASL protocol using a 4-by-3 Hadamard encoding with 3 readout frames, giving 9 decoded timepoints. Segmented 2D radial SPGR readouts were used, each frame comprising 12 excitation pulses (total number of excitations for all readout frames was 108 for sequential vs 36 for time-encoded). The TR was 10 ms, giving a readout frame duration (and therefore temporal resolution) of 120 ms. To match the decoded PLDs between the sequential and time-encoded protocols, the PCASL LD was 360 ms. The first readout frame started immediately after labeling, giving effective PLDs (defined as the time from the end of labeling to the center of each readout frame) of [62, 182, 302, 422, 542, 662, 782, 902, 1022] ms and timepoints (LD+PLD) of [422, 542, 662, 782, 902, 1022, 1142, 1262, 1382] ms.

### 3.2. Flip angle optimization

To fairly compare the sequential and time-encoded PCASL angiography protocols, we optimized the excitation flip angles for each protocol for 2 cases: CFAs and VFAs. In each case, the optimization maximized the minimum theoretical ASL signal across the readout, equivalent to maximizing the profit function:

$$R(N) = \sin(\theta(N)) \cdot \prod_{n=1}^{N-1} \cos(\theta(n)), \quad [2]$$

where we assume a constant supply of labeled blood during the readout, all labeled blood experiences every excitation and there is perfect spoiling of transverse magnetization between excitations. This criterion is equivalent to maximizing the signal at the last excitation for CFAs, meaning Eq. 2 can be simplified to $R(N) = \sin(\theta) \cdot \cos^{N-1}(\theta)$ for CFAs.

To find the optimal CFA for each protocol, a grid search between 1°-60° with a resolution of 0.1° was used. Eq. 2 is maximized for the VFA formula in Eq. 1 when $\theta(N) = 90°$. However, to reduce SAR and the effect of variable slice profiles[16] during the readout, we set $\theta(N) = 30°$. This only leads to a 1% and 4% decrease in the mean signal during the sequential and time-encoded readouts, respectively (see Supporting Information Figure S1).

### 3.3. Signal model simulations

To investigate the advantages of the time-encoding protocol over the sequential, and the VFA scheme over CFA, we used a realistic signal model of labeled blood passing through a single voxel. This model, introduced by Okell et al.,[17] takes into account the LD and PLDs and incorporates RF attenuation, T$_1$ relaxation, and dispersion (characterized by a gamma dispersion kernel). For simplicity, we assume all labeled blood is attenuated by all excitations, which gives the following model for the acquired ASL signal:

$$S(t) = A \cdot R(N') \cdot \frac{s^{1+p \cdot s}}{\Gamma(1+p \cdot s)} \cdot \int_{t-\Delta t-\tau}^{t-\Delta t} e^{-s \cdot t'} \cdot (t')^{p \cdot s} \cdot e^{-\frac{\Delta t + t'}{T_1}} \cdot dt' \quad [3]$$

where $\Gamma(x)$ is the gamma function evaluated at $x$, s is the dispersion kernel sharpness (units of s$^{-1}$), and p is the kernel time-to-peak (units of s). $N' = \left\lceil \frac{t-t_0}{TR} \right\rceil$, where t is the time from the start of PCASL labeling, $t_0$ is the time at the center of the first excitation pulse, $\Delta t$ is the bolus arrival time, $\tau$ is the LD, and TR is the repetition time of the readout (10 ms).

Two cases were considered: (1) no dispersion, with s=10$^4$ s$^{-1}$ and p=0 s, and (2) moderate dispersion, with s=10 s$^{-1}$ and p=0.1 s. To aid comparison with the in vivo data, the moderate dispersion case was down-sampled from a 10 ms temporal resolution to the temporal resolution of the in vivo reconstructed data (120 ms) by taking the mean signal across the 12 excitations in each readout frame.

### 3.4. In vivo comparisons

Six volunteers (3 female, mean age 31±9 years) were recruited and scanned under a technical-development protocol, agreed with local ethics and institutional committees, on a 3T Verio system (Siemens Healthcare, Erlangen, Germany) with a 32-channel receive-only head coil. All scanning occurred during a single scan session for each volunteer. The volunteers were asked to lie still during the scan but were not required to stay awake.

The ASL data were acquired using a 2D SPGR radial readout, similar to that used in Berry et al.[18], to facilitate rapid scans so that all the protocols could be acquired within a single session. A 3-plane localizer was used to place the 70 mm thick, transverse, ASL imaging slab approximately centered at the inferior edge of the thalamus with the superior edge including the corpus callosum (Figure 1D). A single-slab 3D TOF sequence (0.31×0.31×1.3 mm$^3$) was used to place the transverse PCASL labeling plane at the middle of the V3 section of the vertebral arteries.

#### 3.4.1. Acquisition

Two different angiography spatial resolutions were used: (1) a lower in-plane resolution of 1.15×1.15 mm$^2$ to facilitate short scan times, enabling the SNR of all 4 protocols (sequential and time-encoded with both CFA and VFA schemes) to be quantitatively compared, and (2) a higher in-plane resolution of 0.63×0.63 mm$^2$ to qualitatively compare the visibility of small distal vessels for the VFA protocols.

Common imaging parameters were: FOV=220×220 mm, TR=10 ms, 12 excitations per readout frame, fully sampled at k$_{max}$, temporal resolution=120 ms, quadratic RF spoiling[19] (50° increment), no fat suppression. The radial spokes were acquired in sequential angular order within each readout frame, with the sequence acquiring the same spokes for each tag/control/encoding before looping over the next set of spokes until k-space is filled at all timepoints. The spokes were angularly-evenly distributed.

The lower-resolution specific imaging parameters were: matrix=192×192, bandwidth=131 Hz/pixel, TE=5.08 ms, 300 radial spokes, 25 segments (ASL preparations per label/control/encoding), 2:30 min scan time. The sequential protocols used 2 averages to match the scan time of the time-encoded protocols (which require 4 encodings). This data was acquired in 5 of the subjects.

The higher-resolution specific imaging parameters were: matrix=352×352, bandwidth=138 Hz/px, TE=5.27 ms, 552 radial spokes, 46 segments, 4:35 min scan time. For this resolution, the sequential VFA protocol used 1104 separate radial views (92 segments) in order to match the scan time of the time-encoded VFA protocol. This factor of 2 angular oversampling will result in similar noise averaging to the 2-average case and so represents an alternative method for collecting twice as much data for the sequential protocol. This data was acquired in 4 of the subjects.

Balanced PCASL labeling used 500 µs Gaussian RF pulses, 1 ms pulse interval, 20° flip angle, 0.8 mT/m mean gradient, 6 mT/m selective gradient. Background suppression (BGS) used water-suppression-enhanced-through-$T_1$-effects presaturation[20,21] immediately before PCASL labeling and 2 C-FOCI inversion pulses[22,23] (max $B_1^+ \approx 10$ µT due to SAR restrictions, $A_{max}$=20, µ=1.5, and ß=1349.17 rad/s, duration=10.24 ms) interleaved with the PCASL labeling. The PCASL condition was switched from label to control and vice versa after each BGS inversion pulse, to maintain the correct difference signal.[24,25] The inversion pulses were timed to null 2 $T_1$ values (700 and 1400 ms) 100 ms before the first readout excitation using the equation given by Günther et al.[26]

### 3.4.2. Readout under-sampling

When using a 4-by-3 Hadamard encoding, the minimum scan time is twice that of the sequential protocol because 4 encodings must be acquired rather than 2 (label and control). Therefore, to match the time-encoded scan time, we acquired either 2 averages (low-resolution) or twice as many spokes (high-resolution) with the sequential protocol.

However, ASL angiograms are often acquired with only a single average in practice. To investigate the benefit of the time-encoded VFA protocol with the scan time matched to that of 1 average of the sequential protocol, we prospectively angularly under-sampled the readout of the low-resolution time-encoded VFA scan by a factor of ~2 (13 segments compared to 25 for fully-sampled), because the scan time is proportional to the number of segments. This scan (1:18 min), was acquired for all volunteers in the low-resolution comparison and was compared to the first average of the low-resolution sequential VFA scan (1:15 min).

### 3.4.3. Optimized background suppression

For the above experiments, the BGS inversion pulses were timed to null background tissue signal 100 ms before the first excitation. This results in the static tissue signal relaxing during the readout, with physiological noise correspondingly increasing. A more ideal approach would be to null the background tissue signal at some point during the readout, such that physiological noise is optimally reduced throughout the readout.

To explore this idea, we aimed to minimize the maximum acquired background tissue signal during the readout as follows:

$$\min_{t_{null}}[\max(|\mathbf{M_z} \cdot \sin(\boldsymbol{\theta})|)], \qquad [4]$$

where $t_{null}$ is the time when the tissue signal from the 2 target tissue $T_1$ values (700 and 1400 ms) are nulled relative to the first excitation of the readout ($t_{null}<0$ corresponds to after the first excitation and the BGS timing formula ignores the effect of the readout excitations). $\mathbf{M_z}$ is an array of the longitudinal tissue magnetization just before the $n^{th}$ excitation and $\boldsymbol{\theta}$ is an array of the excitation flip angles, where

$$M_z(n) = 1 - [1 - M_z(n-1) \cdot \cos(\theta(n-1))] \cdot e^{-\frac{TR}{T_{1t}}} \qquad [5]$$

for $n \in [2, \ldots, N]$, and

$$M_z(1) = 1 - e^{-\frac{t_{null}}{T_{1t}}}, \qquad [6]$$

where TR=10 ms and $T_{1t}$ is the longitudinal relaxation time of tissue, which we set to the average of white matter and gray matter,[27] $(0.791 \text{ s} + 1.445 \text{ s})/2 = 1.118$ s. This simulation assumes instantaneous RF pulses and perfect spoiling.

The null time was optimized for the sequential and time-encoded VFA protocols using a grid search between -1 and 1 s, with a 1 ms resolution. The inversion pulse timings were constrained to occur during the LD and not to overlap each other. To quantitatively evaluate the effect of this BGS timing optimization on the SNR, in vivo angiography data for both protocols were acquired in 3 of the 5 subjects in the low-resolution comparison.

### 3.4.4. Reconstruction

Offline image reconstruction was performed for the radial angiography data in MATLAB (2021a, The MathWorks, Natick, MA) as implemented in https://github.com/SophieSchau/Accelerated_TEASL.

Before reconstruction, the mean phase difference between matching spokes of the first acquired label/encoding and the following label/control/encoding conditions was subtracted to reduce B$_0$ drift artifacts using $\boldsymbol{k}_m = \boldsymbol{k}_m \cdot e^{-i \cdot \varphi}$, where $\varphi = \angle(\boldsymbol{k}_1^H \cdot \boldsymbol{k}_m)$ and $\boldsymbol{k}_m$ is a single radial k-space spoke across coil channels for the m$^{th}$ label/control/encoding condition. This phase correction assumes that static tissue, which should be identical in each image, is the dominant signal source and that any difference in phase is due to B$_0$ drift. Supporting Information Figure S2 illustrates the reduction in subtraction artifacts after applying this bulk phase correction. To further reduce the impact of B$_0$ drift, the sequence acquired the same spokes for each label/control/encoding condition and any averages before acquiring the next set of spokes, thus minimizing the time between matched k-space data.

After phase correction, density compensation weights were calculated using the fixed-point method[28] and applied. The adjoint of the non-uniform FFT with min-max interpolation[29,30] was used to perform a regridding reconstruction. Coil sensitivities were estimated using the adaptive combine algorithm[31] with a kernel size of $\left\lceil \frac{10 \text{ mm}}{res} \right\rceil$ and a threshold of zero, where *res* is the in-plane resolution. Roemer coil combination[32] was used followed by complex subtraction/decoding. Finally, the magnitude operator was applied to the complex difference images.

### 3.4.5. SNR quantification

A vessel mask was generated for each volunteer using the process outlined in Figure 2 from the combined temporal mean across the 4 scans in the fully-sampled low-resolution comparison. For this process, brain masks were manually drawn for each subject. To calculate the SNR, the mean signal within the vessel mask was divided by the noise SD within the background ROIs at the edges of the image. Because the magnitude operator had been used, the noise SD was estimated as RMS/$\sqrt{2}$ because this is equal to the underlying noise SD, assuming the complex background noise is normally distributed with zero mean and equal SD for the real and imaginary parts. Due to the use of coil sensitivities in the reconstruction, the noise magnitude will vary spatially across the image, but this will occur in a similar manner for all protocols. Significant differences between the calculated SNRs were compared on the subject level using 2-tailed paired t-tests and the Holm-Bonferroni correction for multiple comparisons, using *P*=0.05.

## 4. Results

### 4.1. Flip angle optimization

The CFA flip angle optimization resulted in the following flip angles: sequential=5.5° and time-encoded=9.6°. The minimum flip angle for the VFA formula was sequential=5.45° and time-encoded=9.21°; the maximum flip angle was 30° in both cases, as described above. The flip angles are shown in Figure 3A.

The simulated signal for the constant supply signal model for each case is shown in Figure 3B. The shorter time-encoded readout enabled higher signal to be acquired whichever flip angle scheme was used (mean signal improvement relative to sequential: CFA=74%, VFA=69%). The VFA formula achieved its aim of acquiring constant signal during the readout. It also achieves a higher signal than CFA for most of the readout thanks to the ramp in flip angles (mean signal improvement relative to CFA: sequential=25%, time-encoded=22%).

### 4.2. Signal model simulations

The simulations using the realistic signal model are shown in Figure 3C-E. The signal gain from the larger flip angles possible with time-encoding is also evident with this model. The use of CFAs with time-encoding causes large signal discontinuities between timepoints decoded from different time-encoded blocks, but these discontinuities are smoothed out when using VFAs. Although the signal discontinuities are less obvious at the lower temporal resolution (Figure 3E), the use of VFAs still leads to a clear signal gain compared to CFAs.

When dispersion was included and the temporal resolution matched to the in vivo data (Figure 3E) the relative signal gains were: time-encoded vs sequential mean signal improvement CFA=84% and VFA=69%; VFA vs CFA mean signal improvement sequential=33% and time-encoded=22%. Note, the relative signal gains will depend on the exact model parameters used.

## 4.3. In vivo comparisons

For the low-resolution comparison, Figure 4 shows 3 selected PLDs and the temporal mean for the sequential and time-encoded CFA and VFA protocols for 1 volunteer. The signal gain from using time-encoding vs sequential and VFA vs CFA is apparent at each PLD and in the temporal mean, where smaller vessels can be discerned. Supporting Information Figure S3 shows this data for all subjects at all PLDs.

Figure 5 shows the mean signal at each PLD within 3 vessel ROIs for 1 subject. An early, middle, and late arrival ROI is shown, demonstrating that the signal gains from using time-encoding vs sequential and using VFAs vs CFAs is consistent across the vasculature. Although the bolus arrival time and apparent dispersion differ slightly, the time-encoded CFA signal in Figure 5C has a signal that resembles the temporally-down-sampled simulations in Figure 3E, where the signal peak is lost due to the use of CFAs.

The high-resolution sequential and time-encoded VFA data is shown in Figure 6 for 2 volunteers at 3 selected PLDs along with the temporal means. Similar signal gains can be seen at this resolution with a magnified section of the temporal means demonstrating improved small vessel depiction for the time-encoded scan. Increased subtraction artifacts can be seen at some early PLDs in the time-encoded VFA data due to the larger static tissue signal acquired than in the sequential protocol. Supporting Information Figure S4 shows this data for all subjects at all PLDs.

### 4.3.1. SNR quantification

The quantified SNR for the low-resolution data is shown in Figure 7. The mean SNR was significantly higher in the time-encoded scans compared to the sequential scans (103% increase for CFA and 96% increase for VFA, slightly larger improvements than predicted by simulation). The mean SNR improvement when using VFAs compared to CFAs was also significant (25% for sequential and 21% for time-encoding, similar to the improvements predicted by simulation).

### 4.3.2. Readout under-sampling

Figure 8 shows the 1-average sequential and prospectively angularly under-sampled time-encoded data at 3 selected PLDs and the temporal mean for 1 subject. Despite halving the number of averages for the sequential scan, the vessel visibility is very similar to the 2-average data. This is also the case for the time-encoded data, despite the acquisition being angularly under-sampled by a factor of ~2. The vessels remain more visible for the under-sampled time-encoded data than the 1-average sequential data, demonstrating the time-encoded scan still provides a visibly obvious SNR advantage in a matched scan time to a standard sequential scan. Supporting Information Figure S5 shows this data for all subjects at all PLDs.

The quantitative SNR measures for this comparison are shown in Figure 7. While the sequential SNR only decreased by 6% when a single average was used, the time-encoded SNR decreased by 33% when under-sampled. Nonetheless, the time-encoded SNR was still 39% higher than the sequential SNR.

### 4.3.3. Optimized background suppression

The optimal BGS null times were -87 ms and -217 ms for the sequential and time-encoded VFA protocols, respectively, meaning the static tissue was nulled 87 ms or 217 ms after the first excitation. The null time of the sequential protocol was heavily constrained by the short ASL preparation and the long readout (the BGS inversion pulses could not be placed during the readout). The optimal constrained timings were unchanged for $T_{1t}>0.5$ s, so were robust to the exact value we used.

The simulated static tissue signal at each PLD for the original null time (100 ms) and the optimized null times are shown in Figure 9B. The maximum static tissue signal has been greatly reduced for the time-encoded protocol but has only been slightly reduced for the sequential protocol. The in vivo tissue signal (mean signal within the brain mask) at each PLD for the 3 subjects in Figure 9C exhibits an excellent agreement with the simulations. Figure 9E demonstrates a reduction in the background noise for the time-encoded protocol with the optimized null time, apart from at the 7$^{th}$ PLD. There appeared to be a small increase in background noise for the sequential protocol with optimized null time.

The in vivo SNR for the standard and optimized null times are shown in Figure 9D. The time-encoded SNR increased by 16% (P=0.056) and the sequential SNR decreased by 4% (P=0.15), though neither difference was significant. The in vivo time-encoded temporal means for each subject in Figure 9A highlight areas of reduced physiological noise near the back of the skull, the eyes, and aliased outside the head. Supporting Information Figure S6 shows this data at each PLD.

## 5. Discussion

We have demonstrated that the use of time-encoded PCASL enables the number of excitations in the SPGR readout to be reduced, which facilitates the use of larger flip angles. This approach can approximately double the mean angiographic SNR compared to a time-matched sequential protocol. A VFA scheme that maintains a constant ASL signal throughout the readout results in a smoother signal time-course and increases SNR by a further 21% (145% higher than sequential with CFAs).

By angularly under-sampling the time-encoded acquisition by a factor of ~2, the scan time was reduced to a similar duration to that of a single average of the sequential protocol. Despite this large reduction in k-space sampling, most of the arteries were still clearly visible and there was still a large SNR benefit over the sequential protocol.

Finally, we demonstrated that optimization of the BGS null time can lead to improved static tissue suppression during the entire readout which can reduce the physiological noise present in the angiograms. This is of particular benefit to protocols with short readout durations, which can be achieved with a time-encoded preparation.

## 5.1. Label duration

We chose the relatively short LD of 360 ms in order to match the timings of the sequential and time-encoded protocols. A benefit of short LDs is that arterial inflow can be visualized; if a long LD were used, only the outflow could be visualized. While "inflow-subtraction"[33,34] (where each frame is subtracted from the first frame) can be used to visualize inflow with longer LDs, it increases noise and assumes that all vessels are filled with labeled blood at the start of the readout. Short LDs also reduce scan time; for example, a LD of 1 s would increase the sequential protocol scan time from 2:30 min to 3:34 min (low-resolution) and 4:35 min to 6:33 min (high-resolution). Nonetheless, a long LD could improve the depiction of distal vessels because labeled blood may have reached them before starting to be attenuated by the readout. Furthermore, under severe dispersion, the peak signal can be reduced when using a short LD due to the temporal-smearing of the bolus, while a long LD would still reach a maximum plateau.[35]

## 5.2. Protocol optimization

The protocol timings were chosen in order to directly compare the SNR of matched sequential and time-encoded protocols, similar to the perfusion comparison of Dai et al.[11] In addition to using dynamic angiograms for qualitative assessment of arterial supply, they can provide quantitative measurements of blood volume, transit times, and dispersion parameters when fit with an appropriate kinetic model.[17,36] For this purpose, each protocol may perform best when using a different set of protocol timings, which could be found using a Cramér-Rao lower bound framework to minimize measurement error.[37] This approach could provide insight into the relative merits of each protocol type without constraining the protocol timings to being identical, in a similar way to recent work in quantitative perfusion imaging.[13]

### 5.3. Flip angle optimization

The VFA scheme increased the acquired ASL signal compared to CFAs in simulations and in vivo by more effectively utilizing the available magnetization. Several downsides of this flip angle scheme include higher SAR and increased tissue subtraction artifacts at some PLDs when the BGS is not optimized for the entire readout.

For flip angle optimization, the minimum signal during the readout was maximized, which was equivalent to maximizing the signal at the last excitation for the schemes used here. An alternative approach could use a different VFA scheme to increase the acquired ASL signal throughout the readout, increasing the signal as the tracer travels downstream into smaller vessels. At short PLDs, the labeled blood is concentrated in large vessels where SNR is inherently high, so using reduced flip angles for these PLDs may not be detrimental for vessel visualization. However, this would result in a gradual brightening of the label within a given voxel during the readout and would also cause the signal to oscillate for the time-encoded protocol.

The flip angle optimizations assumed all ASL signal experienced every readout pulse. However, this is unlikely to have been the case with the experimental setup used in this work. For example, 1 volunteer had a 53 mm gap between the imaging volume and the labeling plane. Assuming a constant blood velocity of 40 cm/s and straight vessels perpendicular to the labeling plane and imaging volume, it would take 132.5 ms for labeled blood to reach the image volume, much longer than the 2 ms gap between the end of labeling and the start of the readout. While this likely only has a small visual effect, the resulting signal variations would need to be taken into account for quantification of arterial blood volume.[17]

### 5.4. SNR quantification

The background noise ROIs will incorporate both thermal noise and signal originating from the subject due to the imperfect point spread function, trajectory errors, gridding process, and flow artifacts. When the radial acquisition is angularly under-sampled, these background noise ROIs will also include increased aliased signal. However, this may not greatly affect visualization of the arteries because a circular region with diameter $\frac{FOV}{2}$ around each signal source will only contain a small amount of aliased signal.[38] If most of the arteries with significant signal are within this distance of each other, the under-sampling artifacts will not greatly impact vessel visibility. The SNR calculation for the under-sampled time-encoded data, therefore, will likely be an underestimate of the effective SNR in the local neighbourhood of the vessels and its effective advantage over the 1-average sequential data is expected to be larger than that seen in Figure 8.

## 5.5. Optimized background suppression

The BGS null time for each protocol was optimized to minimize the maximum absolute static tissue signal acquired during the readout. After optimization, the in vivo static tissue was better suppressed across the entire readout. This led to a general reduction in the background noise for the time-encoded protocol, but a small increase in noise for the sequential protocol. Additionally, the noise at the 7th PLD of the time-encoded protocol was increased for all 3 subjects. This noise did not seem to be due to motion because it appeared as a general increase in noise across the entire FOV and it was consistent across subjects (see Supporting Information Figure S6). This consistency suggests it may be due to an interaction between the BGS inversion pulse timings and the encoding pattern of the first time-encoded block (Figure 1C), although further work is required to investigate this.

Overall, the reduced background noise for the BGS optimized time-encoded protocol led to a further 16% increase in SNR on average compared to nulling static tissue before the readout, though this improvement was not significant. Nonetheless, it is widely accepted that reducing static tissue signal reduces physiological noise,[39,40] so despite the limited in vivo demonstration here, we would expect this approach to be generally beneficial.

A more flexible BGS scheme could use inversion pulses during the readout to achieve even better background tissue suppression. Indeed, if this was the case for the 2 BGS inversion pulse method used here, the optimal null time for the sequential protocol would be -685 ms, ~64% of the way through the readout. In contrast, the optimal null time for the time-encoded protocol would be -247 ms (~71% through the readout), only 30 ms later than the null time used in vivo.

## 5.6. Time-encoding

Rather than combining time-encoding with multiple readout frames to generate the temporal information, time-encoding could be the sole source. For example, an 8-by-7 Hadamard encoding matrix with a single readout frame would enable larger flip angles while still yielding 7 PLDs. However, to maintain a similar range of PLDs, the LD would need to be very short (160 ms for PLDs between 62 ms and 1022 ms), potentially leading to a decreased peak signal due to dispersion. A larger encoding matrix also increases sensitivity to subtraction artifacts due to motion. Additionally, to avoid increasing the scan time compared to a single-average sequential protocol, the readout would need to be under-sampled by a factor of 4 rather than 2, as for the 4-by-3 encoding, further increasing aliasing artifacts.

While we kept the LD constant for each time-encoded block for this work, there is no requirement to do so in general.[41] Increasing the LD for the first block will not affect the PLDs or the visualization of inflow, so could be used to further fill the distal vasculature for the longest PLD. However, this approach will increase scan time.

## 5.7. Readout

Although we demonstrated the SNR benefits of time-encoding using radial 2D SPGR, similar SNR benefits would also be expected with other readouts. Balanced SSFP[42] provides greater SNR than SPGR because residual transverse magnetisation is re-used in later TRs rather than being spoiled after each excitation. Time-encoding would enable even larger flip angles than are currently possible with ASL angiography using balanced SSFP.[34] However, it is more challenging to extend balanced SSFP to whole brain 3D readouts due to the technique's sensitivity to off-resonance and the difficulty in achieving a sufficiently good shim across the entire brain at 3T.[34]

## 6. Conclusions

We have demonstrated that a time-encoded preparation can be beneficial for dynamic PCASL angiography by allowing some of the temporal information to be generated from the ASL preparation. This enabled the use of larger flip angles during a shorter SPGR readout which led to an SNR improvement of 103%. Furthermore, a VFA formula was proposed which removed the signal discontinuities present in the time-encoded data when using CFAs and led to a further SNR increase of 21% (145% relative to the sequential CFA protocol). By optimizing the BGS to maintain a low background tissue signal across the readout, the SNR can potentially be increased by a further 16%.

## 7. Acknowledgement

TWO and JGW were supported by a Sir Henry Dale Fellowship jointly funded by the Wellcome Trust and the Royal Society (220204/Z/20/Z). JGW, SSS, and MAC were supported by funding from the EPSRC (EP/L016052/1 and EP/P012361/1). The Wellcome Centre for Integrative Neuroimaging is supported by core funding from the Wellcome Trust (203139/Z/16/Z). For the purpose of open access, the author has applied a CC BY public copyright license to any Author Accepted Manuscript version arising from this submission.

## 8. Data availability statement

In support of Magnetic Resonance in Medicine's reproducible research goal, all of the raw K-space data and all MATLAB (The MathWorks, Natick, MA) code used for reconstruction, simulations, data analysis, and figure generation in this manuscript are available at https://doi.org/10.5281/zenodo.6406741.

## 9. ORCID

Joseph G. Woods https://orcid. org/0000-0002-0329-824X

S. Sophie Schauman https://orcid.org/0000-0002-3744-2553

Mark Chiew https://orcid.org/0000-0001-6272-8783

Michael A. Chappell https://orcid.org/ 0000-0003-1802-4214

Thomas W. Okell https://orcid.org/ 0000-0001-8258-0659

# 11. Figure Captions

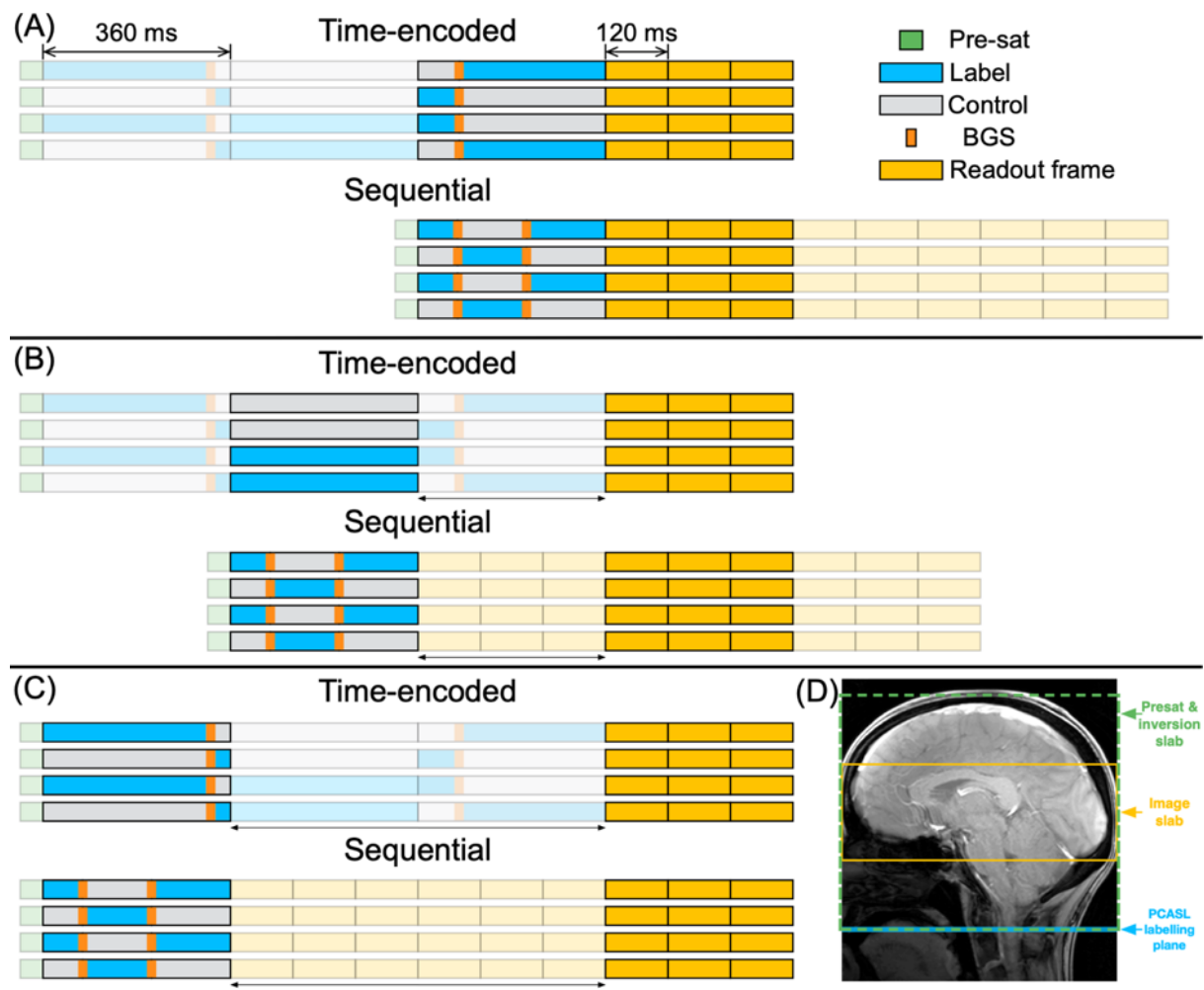

Figure 1: Outline of the protocol timings and in vivo set-up. The use of a 4-by-3 time-encoded preparation with 3 readouts yields 9 dynamic images, which would require 9 readouts in a conventional sequential acquisition. (A-C) demonstrate the equivalent decoded timings between the sequential and time-encoded protocols. For example, in (A), the 3 decoded images from the third time-encoded block give the same information as the first 3 readouts of the sequential acquisition. Note, the PCASL condition switches from label to control and vice versa after each BGS inversion pulse (see text). (D) shows the positioning of the imaging slab, background suppression (presaturation and inversion slabs), and the PCASL labeling plane.

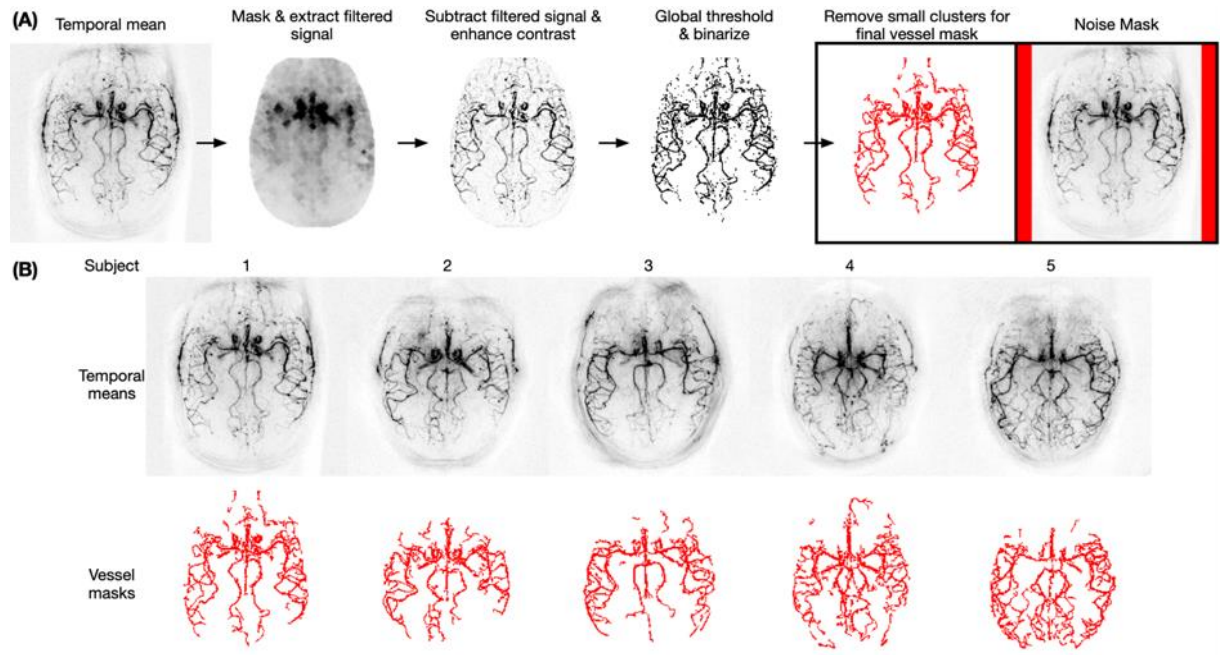

Figure 2: (A) Automated process for creating the vessel masks for the low-resolution comparison with the background noise mask shown in red on the right; (B) the temporal mean images and vessel masks for each subject. The vessel mask used the following heuristic steps: (1) rescale temporal mean to range [0,1]; (2) apply the brain mask and remove the filtered signal calculated using MATLAB's "strel" and "imopen" functions with a 2 voxel disk radius; (3) enhance contrast by saturating the top 1% of voxels; (4) apply a global threshold based on 50% of Otsu's threshold[43] and binarize; (5) remove clusters containing fewer than 10 voxels using MATLAB's "bwareaopen" function.

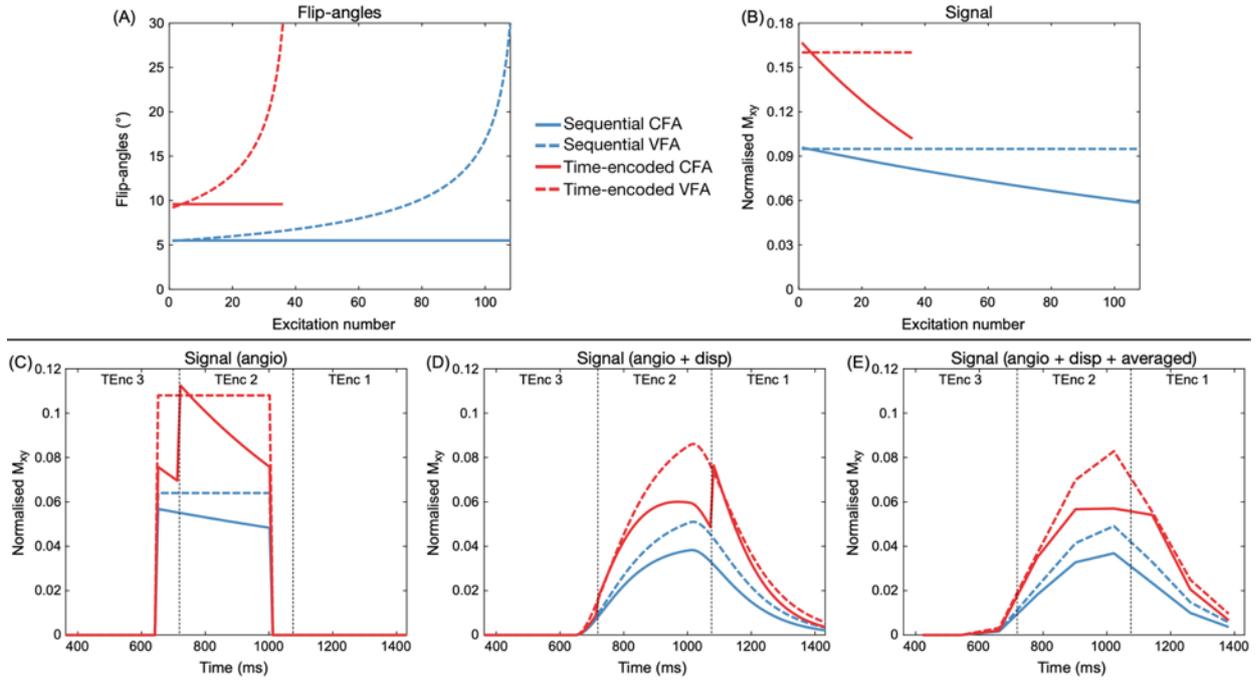

Figure 3: (A) The optimized flip angles for the CFA and VFA schemes for the sequential and time-encoded protocols. (B) The simulated acquired ASL signal for each set of flip angles assuming a constant supply of ASL signal and zero arrival time. The simulated acquired ASL signal using a realistic ASL angiography signal model assuming (C) no dispersion ($s = 10^4$ s$^{-1}$, $p = 0$ s), (D) moderate dispersion ($s=10$ s$^{-1}$, $p=0.1$ s), and (E) moderate dispersion with temporal down-sampling to match the in vivo data. For (C-E), the arrival time was 650 ms and $T_{1b}$ = 1.65 s. TEnc in (C-E) refers to which time-encoded block the data is decoded from.

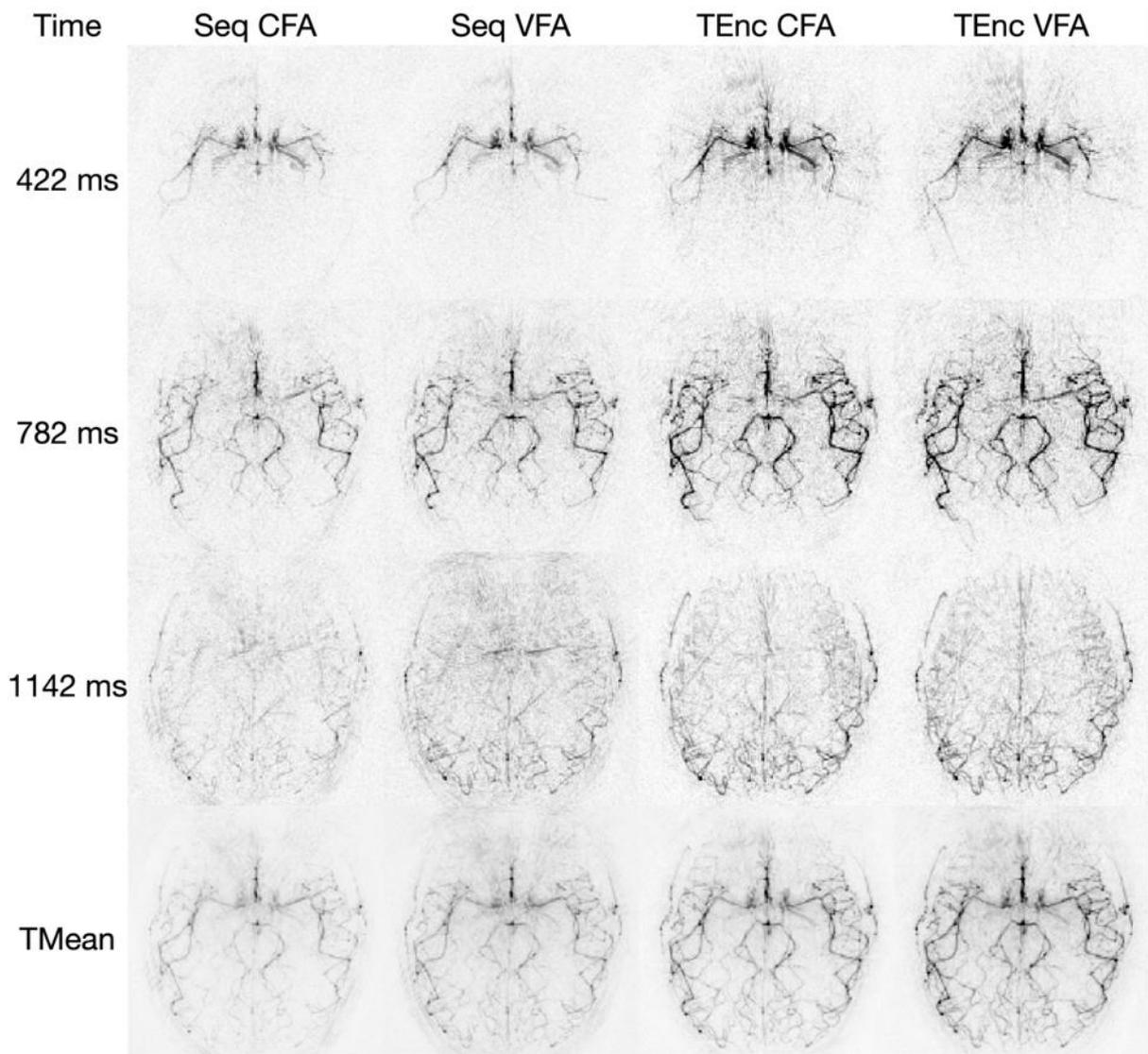

Figure 4: The fully-sampled low-resolution in vivo data for a single subject showing 3 selected PLDs and the temporal mean for each protocol. To aid visualization, 20 voxels around the edges of the FOV have been removed.

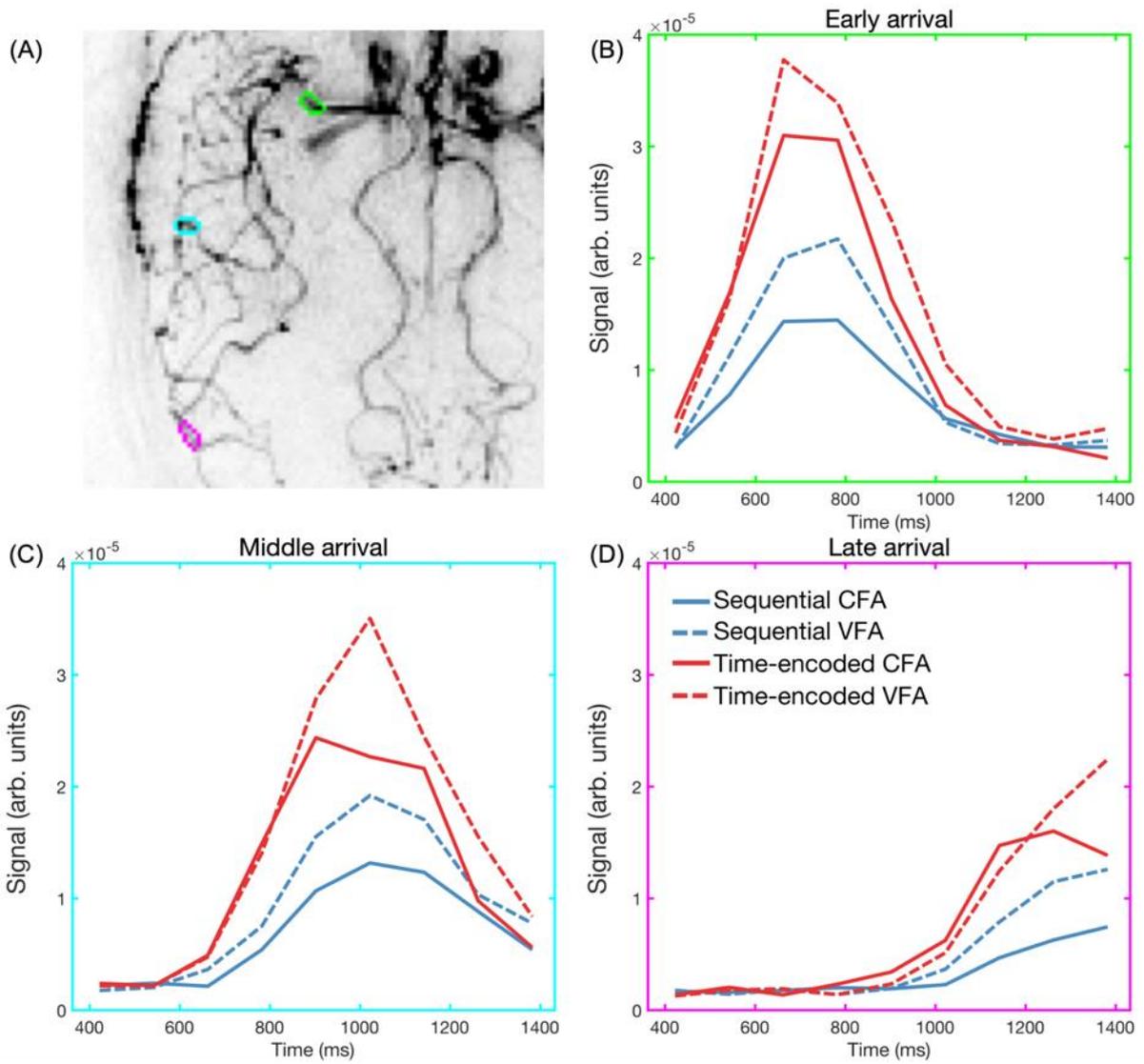

Figure 5: The in vivo signal time-courses across PLDs in 3 vessel ROIs in a single subject. (A) shows the outline of the ROIs overlaid on the subject's temporal mean image. The ROIs show arteries with (A) early, (B) middle, and (C) late signal arrival. Each ROI contained 8 voxels and were selected from the subject's vessel mask shown in Figure 2.

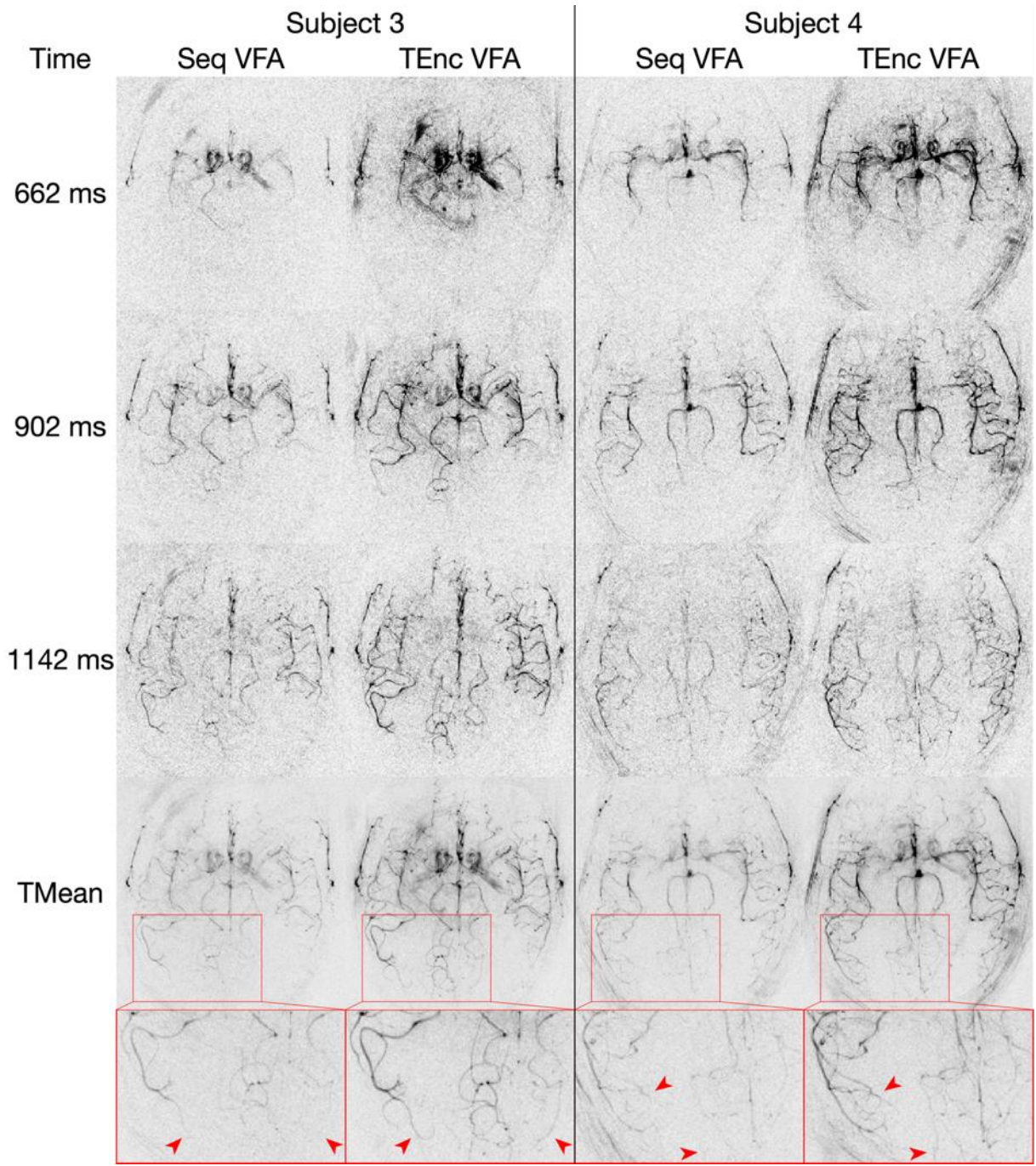

Figure 6: The high-resolution in vivo data for 2 subjects showing 3 selected PLDs and the temporal mean for each protocol. To aid visualization, 40 voxels around the edges of the FOV have been removed. The zoomed section of the temporal mean images illustrate the improved delineation of small distal vessels when using time-encoding (red arrowheads).

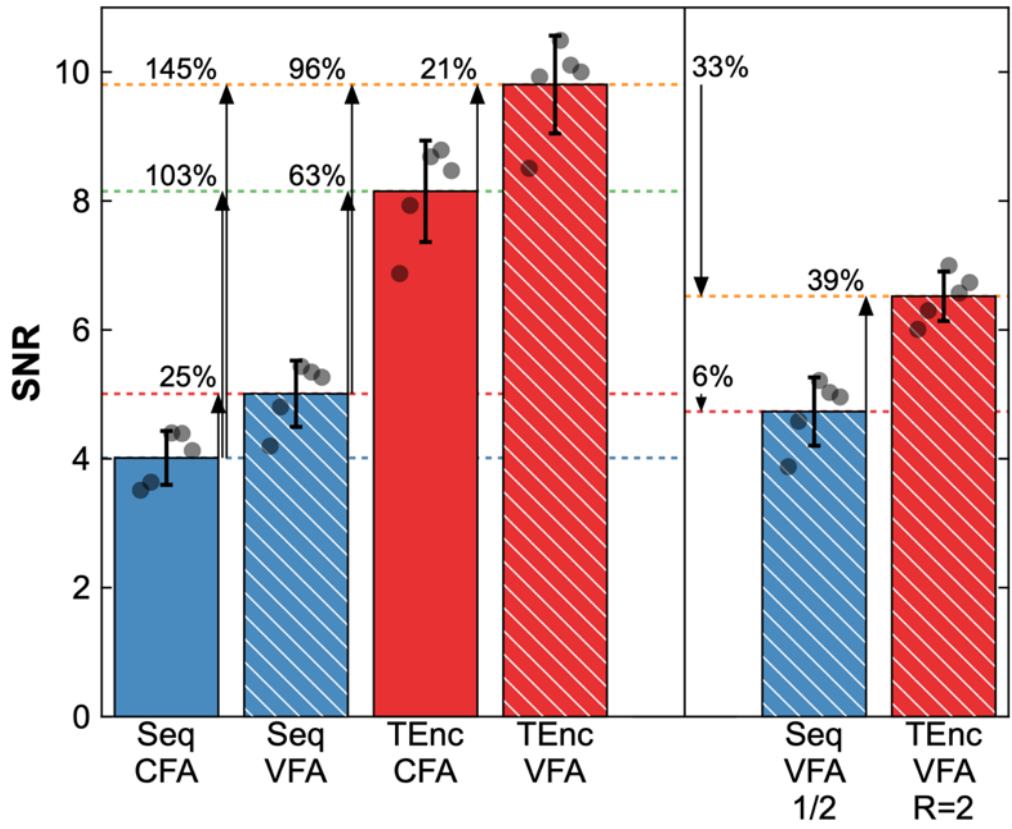

Figure 7: Quantified SNR for the low-resolution data: 2-average/fully-sampled (left) and 1-average/under-sampled (right). The bar graphs and error bars show the mean and SD across subjects. The round markers show the SNR for each subject for each protocol. All differences in the fully-sampled comparison were significant as was the difference in the under-sampled comparison.

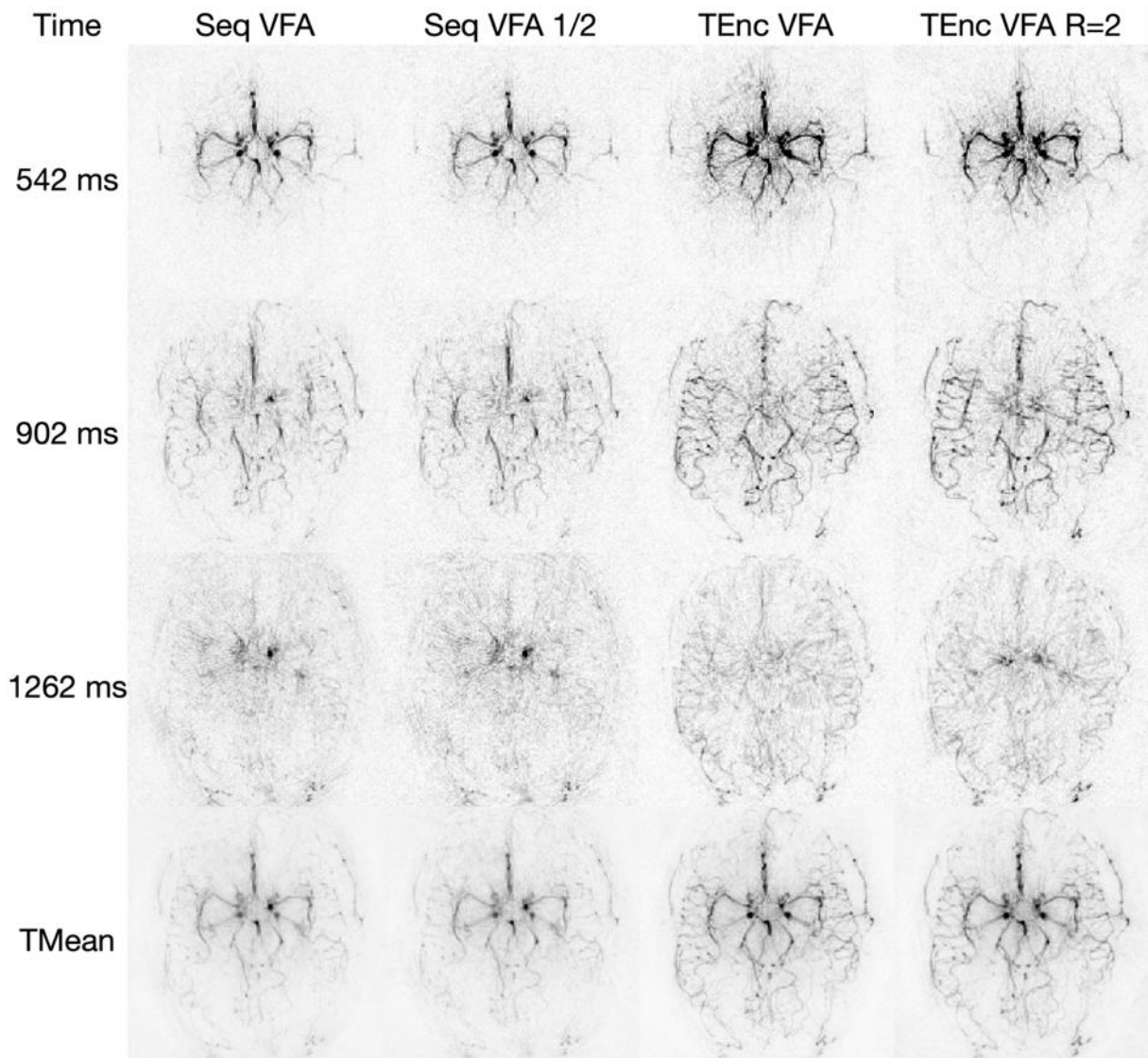

Figure 8: The low-resolution in vivo data for a single subject comparing the 2-average/fully-sampled data to the 1-average/under-sampled sequential and time-encoded data. 3 selected PLDs and the temporal mean are shown for each protocol. To aid visualization, 20 voxels around the edges of the FOV have been removed.

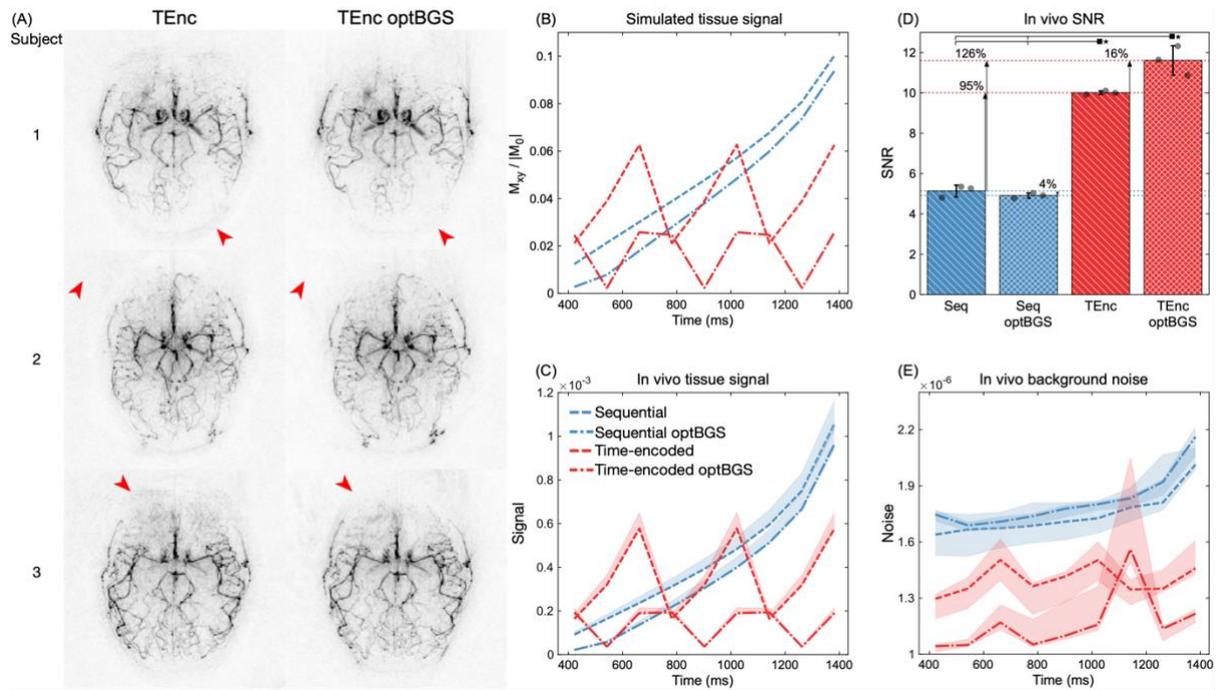

Figure 9: Demonstrating the impact of optimizing the BGS null time. (A) The time-encoded temporal means for each subject using unoptimized and optimized BGS null time, demonstrating an overall reduction in physiological noise; (B) simulated static tissue signal for each protocol (the time-encoded tissue signal for the 3 readouts has been repeated 3 times for visualization); (C) subject-wise in vivo static tissue signal (mean signal within brain mask shown as median [line] and min-max [error envelope] across subjects - time-encoded signal repeated 3 times for visualization); (D) subject-wise in vivo SNR for each protocol using the same vessel masks from Figure 2 (shown as mean [bar] and SD [error bars] across subjects; significant differences relative to TEnc and TEnc optBGS are highlighted with horizontal lines); (E) subject-wise in vivo background noise (shown as median [line] and min-max [error envelope] across subjects).

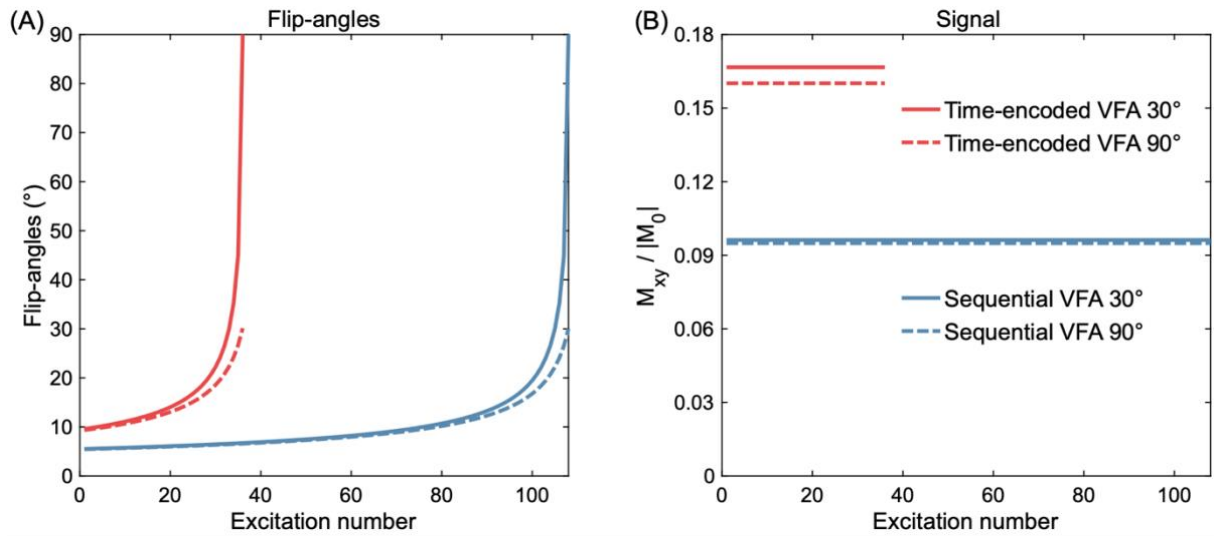

Supporting Information Figure S1: Comparison of the flip angles (A) and simulated acquired ASL signal (B) for the VFA scheme when using either a maximum flip angle of 30° or 90°. The decrease in the simulated acquired signal after reducing the maximum flip angle from 90° to 30° for the sequential and time-encoded protocols was 1% and 4%, respectively. As in Figure 3B, a constant supply of ASL signal and zero arrival time was assumed.

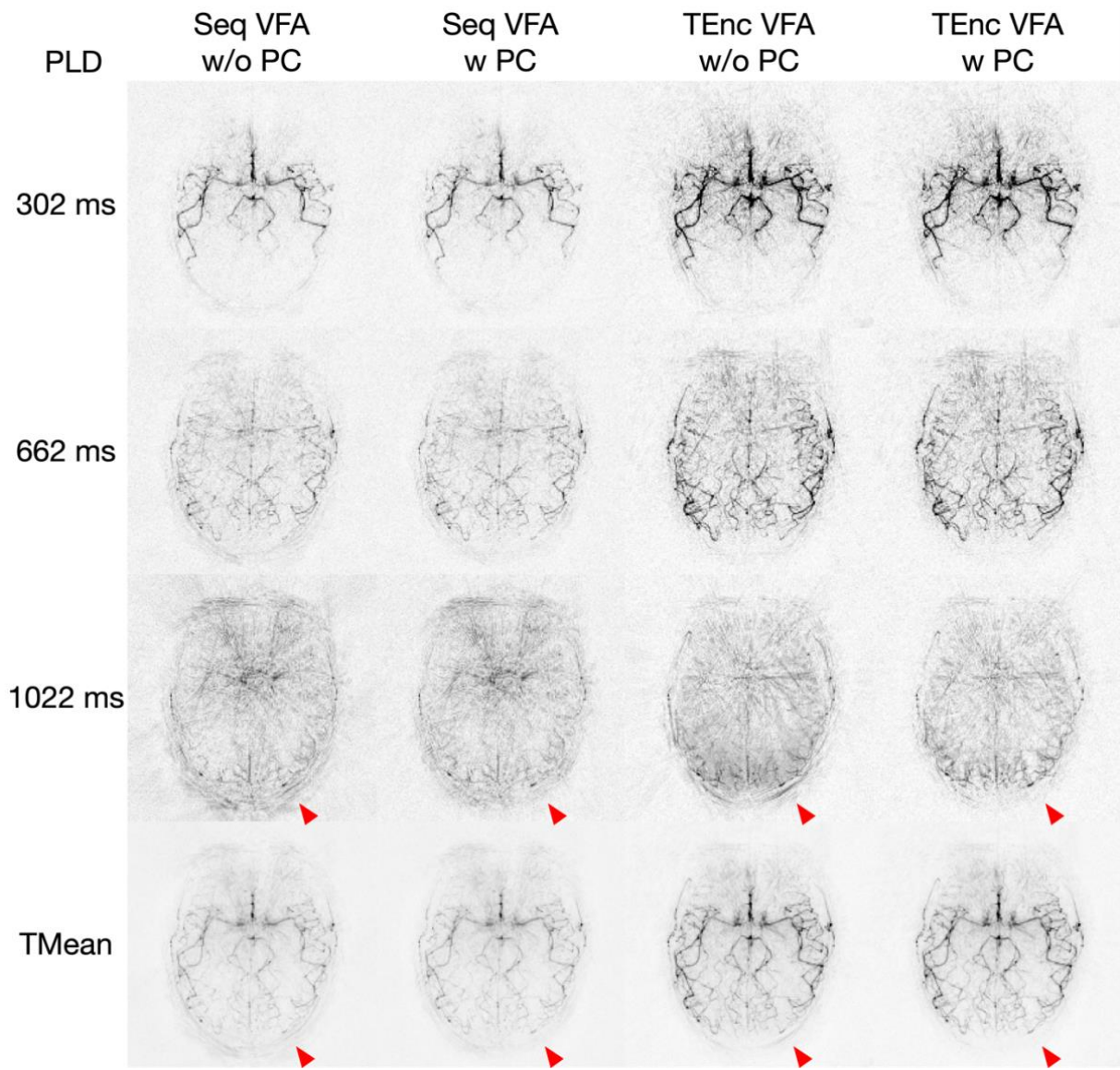

Supporting Information Figure S2: Comparing data reconstructed without and with the phase correction described in the methods section. The use of phase correction helped reduce subtraction errors, most notably at the back of the brain at later time points, when the static tissue signal was larger.

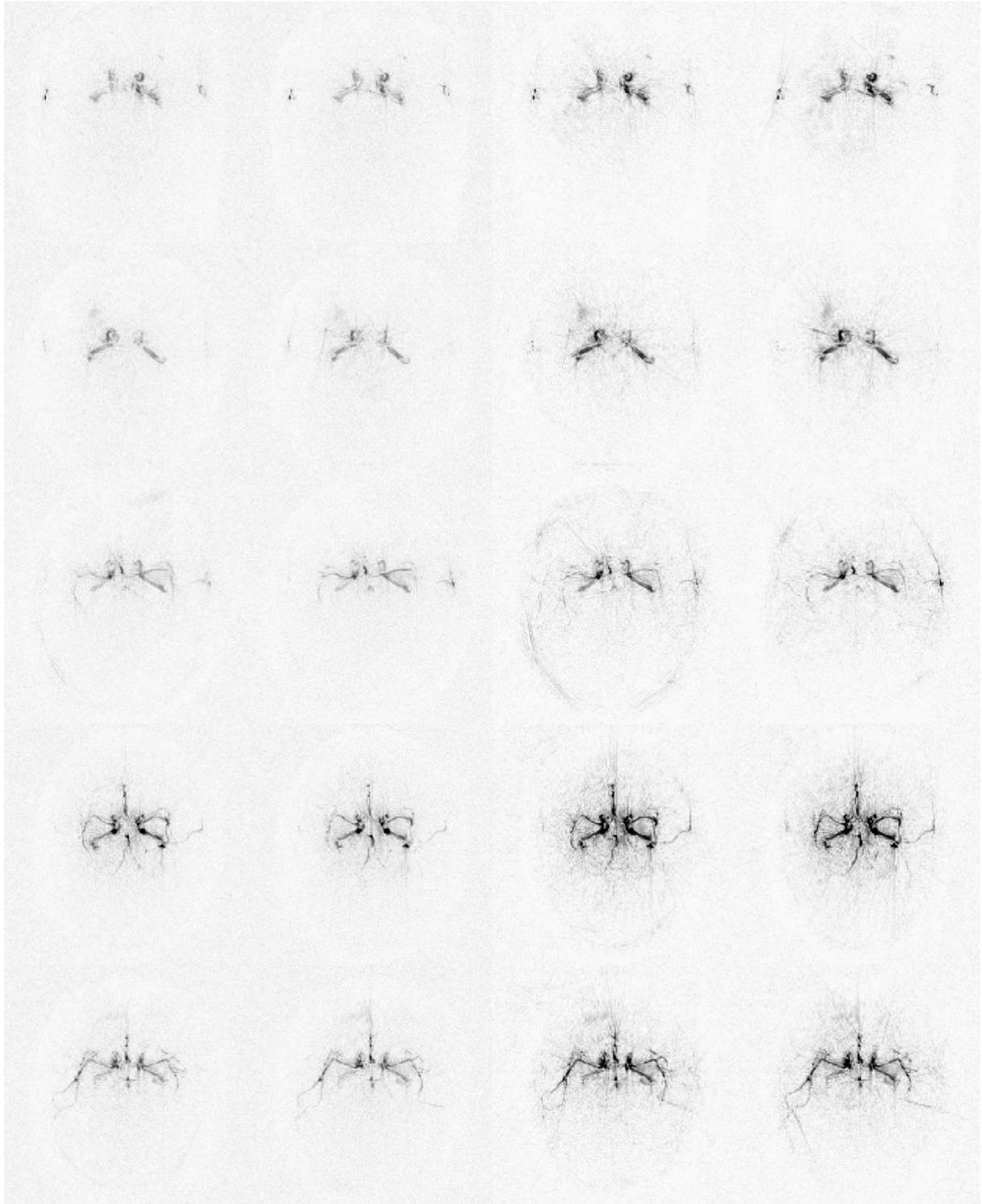

Supporting Information Figure S3: Animation showing the fully-sampled low-resolution data for all 5 subjects at all 9 PLDs. Each row shows a different subject. The columns are (left to right): sequential CFA, sequential VFA, time-encoded CFA, and time-encoded VFA.

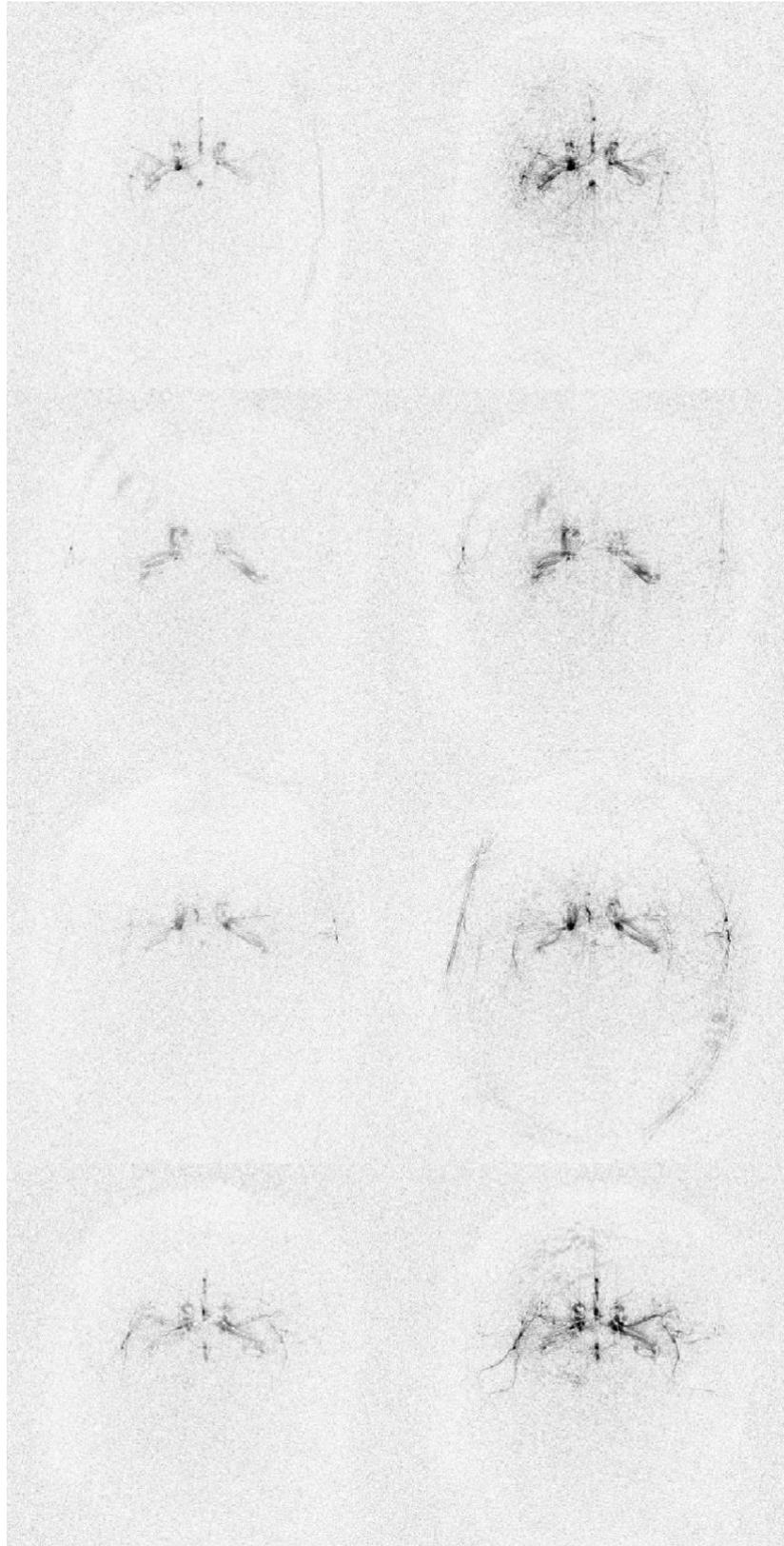

Supporting Information Figure S4: Animation showing the high-resolution data for all 4 subjects at all 9 PLDs. Each row shows a different subject. The columns are (left to right): sequential VFA and time-encoded VFA.

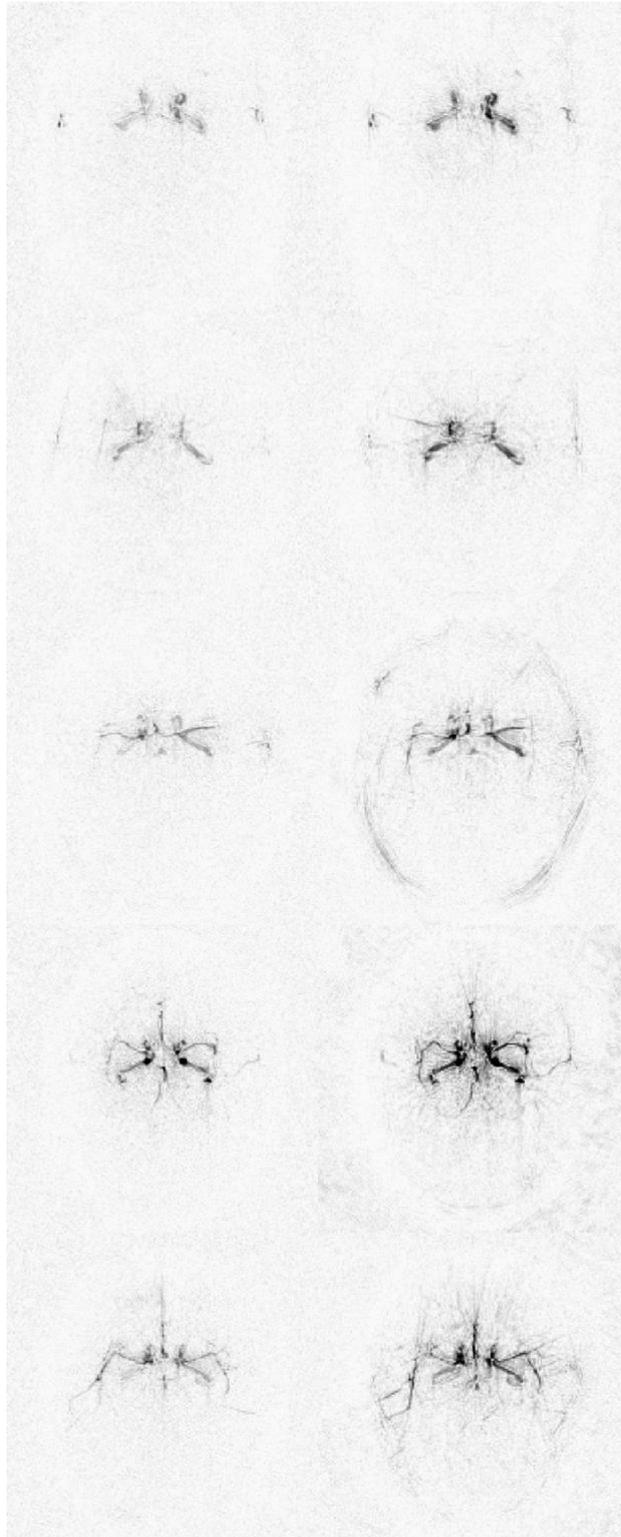

Supporting Information Figure S5: Animation showing the 1-average/under-sampled low-resolution data for all 5 subjects at all 9 PLDs. Each row shows a different subject. The columns are (left to right): 1-average sequential VFA and 2x under-sampled time-encoded VFA.

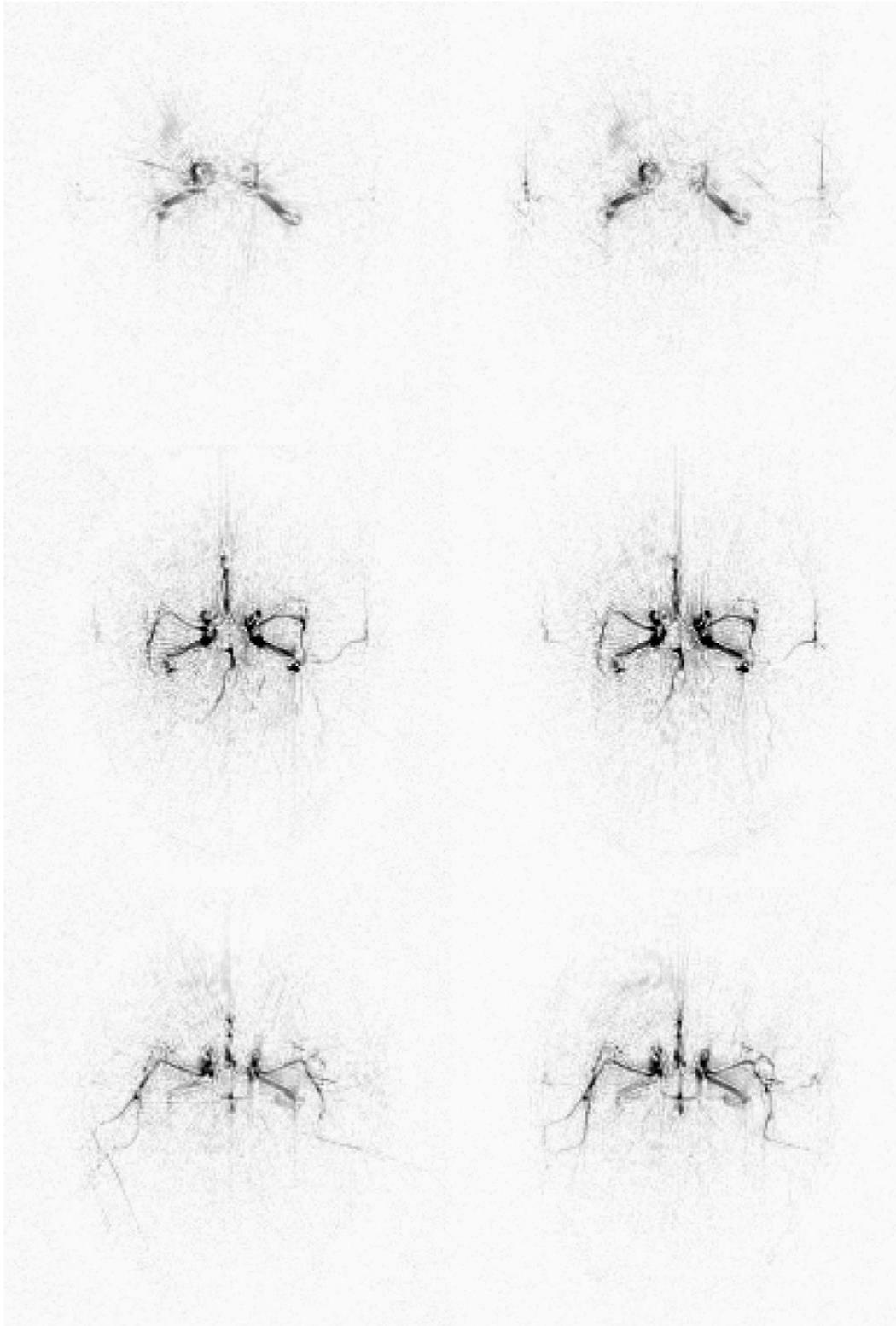

Supporting Information Figure S6: Animation showing the original BGS (left column) and optimized null time BGS (right column) case for the fully-sampled low-resolution time-encoded VFA data for all 3 subjects at all 9 PLDs. Each row shows a different subject. The higher level of background noise in the optimized BGS data at the 7th PLD can be consistently seen for each subject and appears as an increased noise level across the whole FOV.